\documentclass[final,12pt,notitlepage]{article}
\usepackage{amssymb} 
\usepackage{mathtools}
\usepackage{amsthm}
\usepackage[svgnames]{xcolor}
\usepackage[sc,labelsep=period]{caption}
\usepackage[margin=1.2in]{geometry}
\usepackage{graphicx}
\usepackage{csquotes}
\usepackage[authoryear,round,longnamesfirst]{natbib}
\usepackage[inline]{enumitem}
\usepackage[labelfont=rm]{subfig}
\usepackage[bottom,splitrule]{footmisc}
\PassOptionsToPackage{hyperfootnotes=false}{hyperref}
\usepackage[dont-mess-around]{fnpct}
\usepackage[onehalfspacing]{setspace}
\usepackage{titlesec}
\usepackage[colorinlistoftodos,textsize=tiny]{todonotes}
\usepackage{xargs}
\usepackage[normalem]{ulem}
\usepackage{hyperref}
\usepackage[capitalize,nameinlink]{cleveref}

\titleformat*{\subparagraph}{\itshape}
\titlespacing*{\subparagraph}{1.5em}{0.5em}{0.5em}

\hypersetup{
    pdftitle={Communication and Cooperation in Markets},
    pdfauthor={Nageeb Ali and David Miller},
    pdfkeywords={cooperation, community enforcement, social networks,repeated games, network formation},
    colorlinks,
    breaklinks,
    naturalnames,
    hyperfootnotes=false,
    citecolor=DarkRed,
    urlcolor=DarkRed,
    linkcolor=DarkRed
}
\theoremstyle{remark}
\theoremstyle{plain}

\newtheorem{definition}{Definition}

\newtheorem{proposition}{Proposition}
\theoremstyle{definition}

\renewcommand{\epsilon}{\varepsilon}
\providecommand{\indicator}{\mathbb{I}}

\newcommand{\society}{\mathcal{N}}

\newcommand{\buyers}{\society^\mathrm{B}}
\newcommand{\sellers}{\society^\mathrm{S}}

\bibpunct[, ]{(}{)}{;}{a}{}{,}
\newcommand{{\vBS}}{v^\mathrm{S}_{B,S}}
\newcommand{{\vSB}}{v^\mathrm{B}_{B,S}}
\newcommand{{\vBO}}{v^\mathrm{S}_{B,1}}
\newcommand{\lBS}{\lambda_\mathrm{BS}}
\newcommand{\lSS}{\lambda_\mathrm{SS}}
\newcommand{\lBB}{\lambda_\mathrm{BB}}
\newcommand{\qBF}{q^*_\mathrm{BF}}

\newcommand{\VS}{V}

\newcommand{\appendixref}[1]{\hyperref[#1]{Appendix~\ref*{#1}}}
\newcommand{\suppappendixref}[1]{\hyperref[#1]{Supplemental Appendix~\ref*{#1}}}
\MakeOuterQuote{"}

\newcommandx{\nageeb}[2][1=]{\todo[linecolor=blue,backgroundcolor=blue!25,bordercolor=blue,#1]{#2}}
\newcommandx{\david}[2][1=]{\todo[linecolor=green,backgroundcolor=green!25,bordercolor=green,#1]{#2}}
\newcommandx{\important}[2][1=]{\todo[linecolor=red,backgroundcolor=red!25,bordercolor=red,#1]{#2}}

\begin{document}

\title{Communication and Cooperation in Markets}
\author{
	S.\ Nageeb Ali and David A.\ Miller\thanks{
		Ali: Pennsylvania State University. Miller: University of Michigan. We thank Ben Golub for valuable comments and an insightful discussion of this paper. This research was financially supported by NSF grant SES--1127643. 
	}
}

\maketitle

\vspace{1.4\baselineskip}
\begin{abstract}
\noindent Many markets rely on traders truthfully communicating who has cheated in the past and ostracizing those traders from future trade. This paper investigates when truthful communication is incentive compatible. We find that if each side has a myopic incentive to deviate, then communication incentives are satisfied only when the volume of trade is low. By contrast, if only one side has a myopic incentive to deviate, then communication incentives do not constrain the volume of supportable trade. Accordingly, there are strong gains from structuring trade so that one side either moves first or has its cooperation guaranteed by external enforcement.

\end{abstract}
\vfill
\thispagestyle{empty} 
\clearpage

\setcounter{page}{1}
\setstretch{1.25}

\section{Introduction}
In many markets, buyers and sellers can renege on their promises without suffering legal consequences, but defectors are punished by the loss of future business. If a seller trades with many buyers, losing business with a single buyer may not be enough of a threat to deter her from deviating. But if cheating a single buyer results in her losing business with many buyers, then she is more inclined to cooperate. Such schemes, where actions with a single player affect cooperation with others, are at the core of \emph{multilateral} enforcement or "third party" punishment. 
Multilateral enforcement schemes often employ personalized punishment, where traders will work with those who are untainted but sever their ties to those who have deviated in the past. 

For personalized punishment to work, traders need to be able to communicate with each other about their past experiences. Scholars have noted how information-sharing institutions were critical to medieval trade and trust \citep{milgrom/north/weingast:90,greif2006institutions}. Today, online markets rely on ratings and reviews to collect and disseminate information about the behavior of market participants. Credit markets through the ages have benefited from sharing information about borrower histories. 

We view information sharing not as a mechanical process, but as a voluntary choice. If traders are unwilling to communicate truthfully about their past experiences, information will not flow from one relationship to another, making personalized punishment impossible. Thus, we ask: \emph{when do buyers and sellers have a motive to tell the truth to other traders?} 

\paragraph{Model and Results:} We pose this question in a networked market of buyers and sellers, wherein each buyer-seller pair has a long-term trading relationship. Interactions within each relationship are not directly observed by third parties. When a pair interacts, the seller chooses how much quality (or quantity) to deliver to the buyer and the buyer chooses how much to pay the seller. The seller faces an increasing cost function, and thus \emph{may} have an incentive to shirk; the buyer analogously \emph{may} have an incentive to shortchange the seller. In addition to these economic interactions, sellers also randomly meet other sellers, and buyers randomly meet other buyers, just to share information about their past experiences. 

We emphasize that parties  "may" have an incentive to deviate because whether a party \emph{actually} has an incentive to do so depends subtly on the timing of trade. If the buyer and seller act simultaneously, then each party has a myopic incentive to shirk. But if the rules of the marketplace direct the buyer to first make a payment, and the seller then to choose the quantity to trade, then the buyer gains nothing by paying less than the proposed amount, since the seller could then withhold the product. Only the seller has an incentive to shirk. Or the timing might be reversed so that the buyer submits payment only after receipt of the product, in which case only he has an incentive to deviate. Thus, the trading interaction may feature \emph{two-sided} or \emph{one-sided} moral hazard, depending on the timing of trade. 

It is readily apparent that for two-player repeated games,  cooperation is easier to support with one-sided moral hazard than with two-sided moral hazard, because the latter has an additional incentive constraint. We find that the difference is amplified by multilateral enforcement for a new and different reason: it permits some players to share information about others without having to worry about its consequences. We exposit this logic using the class of \emph{permanent ostracism} equilibria, and study how these equilibria perform at a fixed discount rate. We find that permanent ostracism supports substantially more cooperation with one-sided moral hazard than with two-sided moral hazard. 

What is permanent ostracism? It embodies the idea that a trader, Ann, ceases to trade with another, Bob, if she comes to learn that Bob has cheated in the past; however, Ann continues trading with all partners whose reputations are untainted from her perspective. We study these equilibria for two reasons. First, its description matches market behavior where punishments are targeted towards a defector without making the entire market unravel.%
    \footnote{This targeting of punishments towards defectors distinguishes permanent ostracism from contagion \citep{Kandori-Social}, where innocent players shirk on all others once cheated.} 
Second, permanent ostracism offers the simplest scheme in which traders' reputations and records are used to punish or reward them. Thus, it has been the focus of many prior papers, most of which abstract from communication incentives. Our focus is on the effectiveness of these equilibria when traders strategically communicate about who is guilty and innocent.

We study permanent ostracism equilibria at a fixed discount rate, and compare it to two benchmarks. 
The first benchmark is the lower bound of \emph{bilateral enforcement}, which is the most a buyer-seller pair could credibly trade without any third-party punishment. The second benchmark is the upper bound of \emph{naive communication}, which is the highest level of trade achievable if all non-defectors were forced to tell the whole truth, regardless of incentives. Our main results are the following:
\begin{quote}
	\textit{If each buyer-seller pair faces one-sided moral hazard, then permanent ostracism can achieve the benchmark of naive communication. In contrast, if each buyer-seller pair faces two-sided moral hazard, no permanent ostracism equilibrium supports more trade than bilateral enforcement.}
\end{quote}

This result has a clear strategic intuition. In multilateral enforcement, each trader takes on the role of monitoring each other by letting other market participants know if they observe any defections. However, a trader can be trusted only to the extent that he has more to lose in the future than he can gain by defecting. This raises the classic question of "who guards the guardians?" With two-sided moral hazard, each side guards the other, so a trader is unwilling to reveal that she has been cheated because it reduces the degree to which she herself can be trusted. With one-sided moral hazard, first-movers have no myopic incentive to shirk. Hence, they become guards who themselves need not be guarded, and their guardianship secures the cooperation of others.
This difference is sufficiently stark that our negative result holds even when traders obtain verifiable evidence and our positive result obtains even when traders' communication is cheap talk. 

While this result is simple, it elucidates an important point for the design and operation of markets: if word-of-mouth communication is to play a role in supporting trade, there are significant gains from structuring trade (or externally enforcing cooperation) so that one side of the market lacks an incentive to deviate. Doing so amplifies the level of supportable trade, because that side of the market can be relied on to spread news and information.

\paragraph{Related Literature:} The role of word-of-mouth communication in trading relationships has been studied broadly. Many of these studies document the importance of communication \citep{greif:93}, or highlight how its speed and dynamics influence cooperation \citep{raub1990reputation,klein1992promise,ahn:01,Dixit-TradeExpansion}, but abstract from whether traders have the motive to communicate truthfully. A vast literature, surveyed in \cite{tadelis2016reputation}, discusses the role of ratings and feedback in peer-to-peer and online markets. In this literature, the willingness of market participants to disclose their past experiences is often assumed rather than derived.

Our work builds on the study of community enforcement, pioneered by \cite{Kandori-Social}, \cite{Ellison-Random}, \cite{Harrington-Cooperation}, and \cite{OkunoFujiwaraPostlewaite-Social}. One strand of this work envisions that players have "reputational labels" that are updated based on their actions, implicitly assuming that players are sharing information about their past experiences.%
\footnote{Another strand of community enforcement studies folk theorems that obtain in anonymous settings. \cite{Deb-Community} studies general games where players can announce names and authenticate them with their behavior. \cite{DebGonzalez-Community} study how some cooperative outcomes are achievable in some games without communication. Most recently, \cite{DebSugayaWolitzky-Folk} prove a general folk theorem for anonymous random matching. Our work differs from this strand in that we study behavior with a fixed discount rate and focus explicitly on the role of communication in a non-anonymous environment.} 
Our work offers a foundation for these analyses when the interactions involve one-sided moral hazard, and indicates a challenge  when the moral hazard is two-sided. Other recent work studies strategic communication in multilateral enforcement \citep[e.g.][]{LippertSpagnolo-Networks,bowen2012rules,wolitzky2015communication}, illuminating different facets of the problem but do not focus on this distinction between two-sided and one-sided moral hazard for networked markets.
{\cite{BhaskarThomas-Community} study community enforcement with one-sided moral hazard, focusing on centralized, non-strategic information dissemination. \cite{barron2019use} also study communicative incentives with one-sided moral hazard but with a focus on how players may use public communication as a tool for extortion. \cite{SugayaWolitzky-BadApples} study repeated Prisoners' Dilemmas in settings where some players are bad types that never cooperate, and show that communication is essential to cooperation in large groups.}

Finally, this paper relates to our previous work, \cite{AliMiller-Ostracism}, and it is useful to note the differences. Therein we study a society in which each pair of players plays a symmetric repeated Prisoner's Dilemma. In that setting, every connection between players is both a conduit for information and an opportunity for economic behavior. By contrast, in a networked market, an important distinction is between trading links (on which buyers and sellers trade) and communication links (where traders on the same side of the market can talk about their trading partners). While the negative result presented here is rooted in the same strategic forces as that in our prior paper, our main contribution---the positive result for one-sided moral hazard---is fundamentally new. Hence, this contrast between one-sided and two-sided moral hazard, which is our focus here, was absent in our previous work. 

\section{Model\label{Section-Model}}
\subsection{The Setting}
Society comprises buyers $\buyers\equiv\{1,\ldots,{B}\}$ and sellers $\sellers\equiv\{1,\ldots,{S}\}$; let $\mathcal{N} \equiv \buyers \cup \sellers$. A generic buyer ("he") is denoted by~$b$; a generic seller ("she") is denoted by~$s$. The network of relationships in society features both \emph{trading} links and \emph{communication} links. Meetings between buyers and sellers occur on \emph{trading links}: each buyer-seller pair $bs$ meets at random times in $[0,\infty)$ at Poisson rate ${\lBS}>0$. During these meetings, communication and trade occur. \emph{Communication links} involve meetings between players on the same side: each pair of buyers, $bb'$, meets at Poisson rate ${\lBB}>0$, and each pair of sellers, $ss'$, meets at Poisson rate ${\lSS}>0$, to communicate but not to trade. All meeting times are independent across the network. Players share a common discount rate of $r>0$. 

This setting features "local monitoring": if a pair of players is selected to meet at time~$t$, only those two players observe the timing of their meeting and what transpires. Below we describe the extensive form in each of these meetings. 

A buyer-seller interaction spans two stages: first the communication stage, in which they exchange messages, and then the trading stage, in which they trade at a price $p \geq 0$ and a quality $q \in [0,\bar{q}]$. A buyer-buyer interaction or a seller-seller interaction has only the communication phase. We describe the trading stage first.

\paragraph{Trading Stage:}

At this stage, the buyer chooses a payment $p\geq 0$ and the seller chooses the quality of the good, $q\in [0,\bar q]$.%
    \footnote{The upper bound ensures that continuation payoffs remain bounded. We assume the bound is sufficiently high that the constraint never binds.}
When the buyer pays~$p$ and receives quality~$q$, the buyer's payoff is $q-p$ and the seller's payoff is $p-c(q)$. We assume that $c$ is strictly increasing and strictly convex, that $c(0)=c'(0)=0$, and that both $q-c(q)$ and $c(q)/q$ are strictly increasing. 
We consider three distinct extensive forms for the trading stage:
\begin{enumerate}[nolistsep]
	\item \textbf{Simultaneous protocol}: the buyer and seller make their choices simultaneously. 
	\item \textbf{Buyer-first protocol}:  the buyer pays first, and upon receiving payment, the seller chooses how much to deliver to the buyer.
	\item \textbf{Seller-first protocol}: the seller delivers first, and upon receiving delivery, the buyer chooses how much to pay to the seller. 
\end{enumerate}
The simultaneous protocol exhibits two-sided moral hazard, since each partner has a myopic gain from deviating. The latter two protocols exhibits {one-sided moral hazard}, because the party moving first can be immediately punished.

\paragraph{Communication Stage:}

When a pair interacts, along either a trading link or a communication link, they engage in "polite cheap talk" where one of them is randomly selected with probability $\frac{1}{2}$ to speak first, and then after her speech, her partner speaks. The message space enables them to exchange information about which players have deviated. Specifically, the message space is $M \equiv 2^\mathcal{N}$; when player~$i$ sends message $m$ in the equilibria we will construct, the interpretation is that player~$i$ is stating that $m$~is the set of players who are "guilty" because they have deviated. Talk is cheap, so player~$i$ is free to send any message regardless of his or her history.

\subsection{Solution Concept: Permanent Ostracism}

We study a class of equilibria that we call \emph{permanent ostracism} equilibria. We describe the idea intuitively here, relegating formal details to the Appendix. 
A \emph{permanent ostracism} strategy profile is a pure strategy weak perfect Bayesian equilibrium, with an associated system of beliefs, in which each player assesses others as being innocent or guilty. These reputational labels apply to those players that have a myopic gain from shirking, namely all traders in the simultaneous protocol, sellers in the buyer-first protocol, and buyers in the seller-first protocol. In player~$i$'s accounting, all such players begin as innocent. Player~$i$ must deem that partner~$j$ is innocent as long as player~$i$ has not obtained any indication to the contrary. If player~$j$ has a myopic gain from deviating, then player~$i$ should reclassify player~$j$ as guilty if either of the following occur: (1)~player~$j$ fails to exert expected effort or submit expected payment when interacting with player~$i$, or (2)~any player~$k$, when interacting with player~$i$, sends a message $m \ni j$. 
If player~$i$ deems partner~$j$ guilty, she permanently ceases trading with partner~$j$.  
Guilty or innocent, each player communicates truthfully about the behavior of others, but not about herself.
Off the equilibrium path, players' beliefs reflect a correct understanding of the stochastic process governing how information about guilt diffuses through the network given equilibrium behavior. 

\section{A Negative Result for Two-Sided Moral Hazard}\label{Section-TwoSided}

We first describe a negative result for trading games exhibiting two-sided moral hazard. The benchmark for this result is \emph{bilateral enforcement with a simultaneous protocol}. Under bilateral enforcement, the behavior within each buyer-seller relationship depends only on past interactions between them, independently of others. In this benchmark strategy profile, each time they meet, the buyer pays~$p$ and the seller chooses quality~$q$; if either deviates, the pair responds by setting prices and quantities to zero in all future interactions. This grim trigger punishment leads to the incentive constraints
	\begin{align}
		q &\leq q-p +\int_0^\infty e^{-rt} {\lBS}(q-p)\,dt\text{,}
		\tag{Buyer's Bilateral IC}\label{Equation-BuyerBilatIC}
\\
		p &\leq p-c(q) +\int_0^\infty e^{-rt} {\lBS}(p-c(q))\,dt\text{.}
		\tag{Seller's Bilateral IC}\label{Equation-SellerBilatIC}
	\end{align}
Setting both inequalities to bind leads to the highest level of trade supportable by bilateral enforcement, $\underline{q}$, which solves
\begin{align*}
	\frac{c(q)}{q} = \left(\frac{{\lBS}}{r+{\lBS}}\right)^2.
\end{align*} 
While multilateral enforcement could, in principle, improve upon bilateral enforcement, we show that no permanent ostracism equilibrium can do better. 

\begin{proposition}
	\label{Prop-PermOst}
	With two-sided moral hazard, in every permanent ostracism equilibrium, the level of trade never exceeds $\underline{q}$ in any equilibrium path history.
\end{proposition}

Here is the logic (the proof is in the Supplementary Appendix): if the level of trade exceeds~$\underline{q}$, then Ann trusts Bob if she believes others are innocent and available to punish Bob. But if Bob knows that others have deviated, and divulging this information to Alice will make her trust him less, he has no incentive to do so. He is better off concealing it and shirking on Ann.
Since all he needs to do is conceal, the conclusion holds even if Bob has verifiable evidence that others have deviated.

\section{Positive Results for One-Sided Moral Hazard}
This section presents our main contribution: communication is incentive compatible at high levels of trade if the stage game exhibits one-sided moral hazard. For concreteness, we consider a buyer-first protocol; a similar result obtains for a seller-first protocol. 

Before proving our result, we first describe a \emph{naive communication} benchmark that would be relevant if innocent traders were \textit{forced} to communicate truthfully in all of their interactions. A player who is guilty will be punished by every partner he meets who deems him guilty. The best a guilty player can hope is to cheat every unsuspecting partner he meets. This logic leads to the incentive constraints:
	\begin{align}
		0 &\leq q-p + {S} \int_0^\infty e^{-rt} {\lBS}(q-p)\,dt\text{,}
		\tag{Buyer's IC}
		\label{Equation-BuyerBFirstIC}
		\\
		p + (B-1) p \,{\vBS} &\leq p-c(q) + {B} \int_0^\infty e^{-rt} {\lBS} (p-c(q)) \,dt \text{.}
		\tag{Seller's IC}
		\label{Equation-SellerBFirstIC}
	\end{align}
The \ref{Equation-BuyerBFirstIC} reflects that  so long as he is gaining from trade ($q\geq p$), he must be better off from cooperating both myopically and in terms of discounted continuation value.	

By contrast, the \ref{Equation-SellerBFirstIC} reflects that the seller has a myopic incentive to deviate whenever $c(q)>0$, and her motive for not doing so is to maintain cooperation with future buyers. 
In \ref{Equation-SellerBFirstIC}, $\vBS$ describes the (discounted) probability that when the seller next meets a given buyer, he has not yet learned of her guilt; naturally, $\vBS$ depends on the rate at which information about a guilty seller diffuses in the market.\footnote{We show in \cref{Footnote-Viscosity} how to compute ${\vBS}$ recursively from primitives.} 
There are several ways for information to diffuse: seller~$i$ may cheat on another unsuspecting buyer, or a buyer or seller who knows seller~$i$ is guilty may pass on that information.%
\footnote{This benchmark uses both buyers and sellers as conduits of information about sellers' behavior, so as to maximize the level of cooperation. Later, we discuss equilibria in which only buyers communicate.} 
It is only in the first case that seller~$i$ accrues payoffs from others learning about her guilt, and ${\vBS}$ captures the discounted probability of her obtaining a payoff from these future defections.
Accordingly, the term $(B-1)p{\vBS}$ measures the discounted value of seller~$i$'s future opportunities to cheat other buyers in the future, before they have learned that she is guilty.
	
Setting these constraints to bind and simplifying yields the maximum level of cooperation with naive communication under a buyer-first protocol, ${\qBF}$, which solves
	\begin{align}
		\frac{c(q)}{q} = \frac{{B} {\lBS} - (B-1) r {\vBS}}{r + {B} {\lBS}}\text{.}
		\label{Equation-BFirstCost}
	\end{align}
This is our benchmark for high cooperation. All of the "social collateral" here is put on the side of the sellers: because buyers have no myopic incentive to deviate, they can be deterred from deviating within the stage game and don't need to be rewarded or punished through continuation play. By contrast, the seller is deterred by multilateral enforcement, where she puts several relationships at risk if she cheats in any one relationship. Thus, she is willing to produce higher quality in each relationship than under bilateral enforcement.%
	\footnote{
		This holds for both bilateral enforcement under a simultaneous protocol, which yields quality~$\underline{q}$, and for bilateral enforcement under a seller-first protocol, which yields quality that solves $\frac{c(q)}{q} = \frac{{\lBS}}{r + {\lBS}}$.
	}

But this benchmark is an unsatisfactory description of behavior because it assumes that innocent players are simply forced to reveal the truth. Our positive result is that this benchmark is attainable, even if players can strategically choose what to disclose: we construct a permanent ostracism equilibrium that attains this naive communication benchmark~${\qBF}$.

In this equilibrium, only sellers can be considered guilty; because a buyer has no incentive to deviate, there is no reason to ostracize him. When a buyer meets a seller he believes is innocent, he first pays~$p$ (regardless of whether he has deviated in the past). Then the seller produces quality~$q$ if she is innocent and price~$p$ was paid; she produces zero quality otherwise. If the seller deviates in the trading stage, she becomes guilty, and thereafter aims to shirk on all future buyers too. Buyers, in turn, aim to ostracize guilty sellers while continuing to trade at equilibrium-path levels with innocent sellers.

Players communicate truthfully, with the exception that each seller never communicates about herself.
Consequently, an innocent seller's incentive constraint is \ref{Equation-SellerBFirstIC} from above, enabling cooperation at price and quality $p = q = {\qBF}$ both on and off the equilibrium path. We show that a guilty seller has an incentive to cheat on every buyer once she has cheated on anyone.\footnote{This argument is similar to how in contagion, contagion phase incentive constraints are necessarily satisfied if cooperation phase incentive constraints hold with equality, as in \cite{Ellison-Random}.}
In this equilibrium, buyers have no incentive to deviate at the trading stage. Moreover, neither sellers nor buyers have any incentive to lie. 
These observations lead to our main result:
\begin{proposition}\label{Prop-MainResult}
	With a buyer-first protocol, there exists a permanent ostracism equilibrium that attains the naive communication benchmark.
\end{proposition}

The logic is that with a buyer-first protocol, an innocent seller meeting a buyer does not care whether that buyer has valuable relationships with other sellers---the buyer can be trusted to pay in any case. Thus, we can make a buyer's payoff from each interaction independent of his message, which makes him willing to truthfully reveal whether other sellers have defected.%
\footnote{We note that while in our construction, the buyer obtains zero payoffs in each relationship, this is unnecessary for communication incentives. All that is needed is that his payoff in any interaction with an innocent or guilty seller is independent of his report.} 
Because buyers themselves are guarded---by having to move first---their communication guards the cooperation of others. 

We describe several attractive properties of this equilibrium. First, consistent with the theme of personalized punishment, it ensures that a few rotten apples do not spoil collective cooperation. Second, the equilibrium does not require coordination on a common calendar or start date.%
\footnote{\cite{clark2029steady} also view this as an attractive property for multilateral enforcement and develop schemes in anonymous environments that do not require such coordination.} 
Third, it may be that an intricate construction is needed to increase the average level of trade beyond this equilibrium; standard constructions such as contagion equilibria \citep{Kandori-Social,Ellison-Random} cannot. 

Before we prove this result, we make two further remarks. First, the same approach also works for a seller-first protocol: a seller who deviates can be punished immediately by the buyer, while buyers need to be disciplined by permanent ostracism. A similar line of reasoning leads to an expression comparable to \eqref{Equation-BFirstCost} for the highest level of supportable trade. 
One could investigate which of seller-first and buyer-first protocols is better; such a comparison subtly depends on the model's primitives.
	
Second, our model assumes that sellers impose no externalities on each other. In many settings, sellers may generate externalities, and hence cannot be trusted to report truthfully about each other. Buyers can still be relied on, however, so one could construct an analogous equilibrium that uses only buyer-to-buyer communication. Because information would spread more slowly, the supportable level of trade would be lower than $\qBF$.%

\begin{proof}[Proof of \cref{Prop-MainResult}]
	We pair the strategy profile described above with a system of beliefs in which buyers believe that sellers are innocent until revealed to be guilty, either by their own actions or via communication from other players.  
	We consider the incentives of buyers, innocent sellers, and guilty sellers in turn. 
	
    \subparagraph{Buyer Incentives:}
    At every interaction, the equilibrium prescribes that a buyer communicate truthfully; following communication, the buyer pays $p=\qBF$ to a seller he deems innocent, and pays zero to a seller he deems guilty. Because the buyer's payment does not affect continuation play (with either the same seller or any other), his incentive to pay pertains only to the current period. When facing an innocent seller, he has no myopic incentive to deviate, because any deviation leads the seller to subsequently deliver zero quality. When he meets a seller he deems guilty, he expects the seller to deliver zero quality regardless of his payment, so there is no incentive to pay more than zero.  Finally, the buyer does not gain by deviating in the communication stage of any meeting, because, regardless of his message, he obtains the same payoff in every interaction.
	
\subparagraph{Innocent-Seller Incentives:}
   Both on and off the equilibrium path, an innocent seller's incentive constraint to produce quality $q=\qBF$ is \ref{Equation-SellerBFirstIC}, which is satisfied with equality: in each case, she expects all buyers to continue communicating truthfully and cooperating with her regardless of the guilt or innocence of any other seller and regardless of the past deviations of any buyer. Similarly she expects all buyers and all other sellers, guilty or innocent, to communicate truthfully about her guilt or innocence. Finally, there is no contingency where a deviation from truthful communication improves her payoff.
   
 \subparagraph{Guilty-Seller Incentives:}
   Equilibrium strategies prescribe that a guilty seller communicates truthfully about other sellers and produces zero quality. A guilty seller has no incentive to misreport about other players, because her report doesn't affect what happens in her current or subsequent interactions. 
   
   We now prove that a guilty seller finds it incentive compatible to produce zero quality. Suppose, without loss of generality, that seller $s'$ and buyer $b'$ meet at time $0$, and seller $s'$ deviates and produces zero quality. Suppose that at time $t$, the seller meets buyer $b''$. If seller $s'$ already knows that buyer $b''$ deems her to be guilty---either because she has already deviated in a prior meeting with $b''$ or because $b''$ told her so in the communication stage---then she has no incentive to deviate to higher quality. If instead $b''$ deems her innocent and pays the equilibrium-path price $p=\qBF$, then the seller could deviate once to choosing quality~${\qBF}$ (so as to delay buyer~1 from learning of her guilt). A sufficient condition for the seller to choose zero quality rather than deviate to ${\qBF}$ is if, for every $k_b\geq 1$ and $k_s\geq 1$,
   \begin{align}
		\qBF + \qBF  {\VS}(k_b+1, k_s) 
		&\geq \qBF - c(\qBF) + \qBF {\VS}(k_b, k_s)\text{,}
		\label{Equation-SellerBFOffPathIC}
	\end{align}
	where $\qBF {\VS}(k_b, k_s)$ is her expected continuation payoff when $k_b$~buyers and $k_s$~sellers (including herself) deem her guilty.%
	\footnote{\label{Footnote-Viscosity}%
	The value of ${\VS}(k_b,k_s)$ is computed from a recursive system of equations: for $k_b\in \{1,\ldots,B\}$ and $k_s\in \{0,\ldots,{S}\}$, let 
	\begin{align}\begin{aligned}
		{\VS}(k_b,k_s) &= \int_0^\infty \left( 
			\begin{aligned}
	 			&e^{-rt} e^{-\left( {\lBB} k_b ({B}-k_b) + {\lSS} (k_s-1) ({S}-k_s) + {\lBS} k_b ({S}-k_s) + {\lBS} k_s ({B}-k_b)\right) t}
				\\ &\cdot \left(
					\begin{aligned}
						&{\lBB} k_b({B}-k_b) {\VS}(k_b+1,k_s) + {\lSS} (k_s-1) ({S}-k_s) {\VS}(k_b,k_s+1) 
						\\
						&+ {\lBS} k_b ({S}-k_s) {\VS}(k_b,k_s+1) + {\lBS} k_s (B-k_b) {\VS}(k_b+1,k_s) + {\lBS} (B-k_b)
					\end{aligned}
				\right)
	 		\end{aligned}
	 	\right) dt \text{,}
	\label{Equation-ViscositySystem}
	\end{aligned}
	\end{align}
where ${\VS}(B,k_s)=0$ for each $k_s$. Finally, let $v^S_{B,S}\equiv {\VS}(1,1)/(B-1)$. The recursive equation \eqref{Equation-ViscositySystem} documents how the seller waits until the next interaction that diffuses information about her guilt, which happens only when an "informed" trader meets an "uninformed" trader. The guilty seller reaps surplus only when she meets an uniformed buyer, which occurs with density $\lBS (B-k_b)$. 
	}
	Although seller~$s$ is uncertain about $(k_b,k_s)$, if \eqref{Equation-SellerBFOffPathIC} holds pointwise for every realization of $k_b\in \{1,\ldots,B\}$ and $k_s\in\{1,\ldots,S\}$, then it is sequentially rational for her to produce zero quality. 
	Observe that \eqref{Equation-SellerBFOffPathIC} can be re-arranged to
	\begin{align*}
	    	\frac{c({\qBF})}{\qBF}\geq {\VS}(k_b,k_s)-{\VS}(k_b+1,k_s).
	\end{align*}
We verify that this inequality is satisfied for every $k_b\geq 1$ and $k_s\geq 1$. Because $\qBF$ binds the equilibrium path incentive constraint, a seller is just indifferent between the equilibrium path and producing zero quality in every trading stage. Let $\qBF {\VS}(0,1)$ be her continuation payoff if she has been on the equilibrium path until now but plans to shirk on the next buyer she meets. Then her binding \ref{Equation-SellerBFirstIC} can be re-written as
\begin{align}\label{Eq-BindingIC}
\qBF+\qBF {\VS}(1,1)=\qBF-c(\qBF)+\qBF {\VS}(0,1). 
\end{align}
Re-arranging \eqref{Eq-BindingIC} implies that 
		\begin{align*}
			\frac{c({\qBF})}{\qBF}
			&=  
			 {\VS}(0,1) - {\VS}(1, 1)
			\text{.}
		\end{align*}
Thus, \eqref{Equation-SellerBFOffPathIC} is satisfied if
	\begin{align}\label{eq:Contagion phase claim}
	{\VS}(k_b,k_s) - {\VS}(k_b+1,k_s) \leq {\VS}(0,1) - {\VS}(1,1).
	\end{align}
 We prove that \eqref{eq:Contagion phase claim} is satisfied for all $k_b = 0, \ldots, B$ and $k_s = 1, \ldots, S$,, adapting the argument of Lemma 1 of \cite{Ellison-Random}. %
 We consider every sequence of link recognitions in which no two links meet simultaneously, and then take expectations over them. Let $\xi = (\tau_z,\ell_z)_{z=1}^\infty$ be a  sequence of link recognitions that take place in time span $[0,\infty)$, where $\left(\tau_z\right)_{z=1}^{\infty}$ is the ordered list of link recognition times and $\left(\ell_z\right)_{z=1}^{\infty}$ is the list of links in their order of recognition. (We define a link $\ell$ between player~$i$ and player~$j$ as a set $\{i,j\}$. But with some abuse of notation, we also say that $\ell \in A\times B$ if either $(i,j) \in A\times B $ or $(j,i) \in A\times B$.)
 
	Let $K^0 = (K_b^0,K_s^0)$ be the initial "$s$-state"---the sets of buyers and sellers (respectively) who deem seller~$s$ guilty, at a start time normalized to zero. Then, if the sequence of link realizations is~$\xi$, the $s$-state immediately following the interaction at time $\tau_z$ is
		\begin{align*}
			\begin{aligned}
				&\bigl(\kappa_b^z(K^0,\xi),\kappa_s^z(K^0,\xi)\bigr) 
				\\
				&\quad= 
				\begin{cases}
					(K_b^0,K_s^0) &\text{if $z=0$,}
					\\
					\bigl(\kappa_b^{z-1} \cup (\ell_z \cap \buyers), \kappa^{z-1}_s\bigr) &\text{if $z>0$ and $\ell_z \in (\buyers \setminus \kappa_b^{z-1}) \times (\kappa_b^{z-1} \cup \kappa^{z-1}_s)$,} 
					\\
					\bigl(\kappa_b^{z-1},\kappa_s^{z-1} \cup (\ell_z \cap \sellers)\bigr) &\text{if $z>0$ and $\ell_z \in (\sellers \setminus \kappa_s^{z-1}) \times \bigl((\kappa_b^{z-1} \cup \kappa^{z-1}_s) \setminus \{s\}\bigr)$,}
					\\
			    	\bigl(\kappa_b^{z-1},\kappa_s^{z-1}\bigr) &\text{otherwise,}
				\end{cases}
			\end{aligned}
		\end{align*}
	where, with some abuse of notation, we write $\bigl(\kappa_b^{z-1},\kappa_s^{z-1}\bigr)$ for $\bigl(\kappa_b^{z-1}(K^0,\xi),\kappa_s^{z-1}(K^0,\xi)\bigr)$.
	Define $\tilde{V}\bigl(K_b^0,K_s^0\bigm|\xi\bigr)$ to be the equilibrium continuation payoff of seller~$s$ when the initial $s$-state is $K^0 = (K_b^0,K_s^0)$ and the sequence of link recognitions is~$\xi$. 
	The change in seller~$s$'s continuation payoff when one more buyer~$j$ deems her guilty at the outset, for any $j \in \buyers$,
		\begin{align}\label{eq:Contagion phase mess}
			\begin{aligned}
				&\tilde{V}\bigl(K_b^0,K_s^0 \bigm|\xi\bigr) - \tilde{V}\bigl(K_b^0 \cup \{j\}, K_s^0 \bigm|\xi\bigr)
			\\
				&\quad= \sum_{z=1}^{\infty} e^{-r\tau_z}
					\sum_{b\in \buyers} \qBF
					\indicator \bigl( \ell_z = \{s,b\} \text{ and }
					b \in \kappa_b^{z-1}((K_b^0 \cup \{j\}, K_s^0),\xi) \setminus \kappa_b^{z-1}((K_b^0, K_s^0),\xi) \bigr)
			\\
				&\quad\leq \sum_{z=1}^{\infty} e^{-r\tau_z}
					\sum_{b \in \buyers} \qBF
					\indicator \bigl( \ell_z = \{s,b\} \text{ and }
					b \in \kappa_b^{z-1}((\{j\},\{s\}),\xi) \setminus \kappa_b^{z-1}((\emptyset,\{s\}),\xi) \bigr)
			\\
				&\quad= \tilde{V}\bigl(\emptyset,\{s\} \bigm|\xi\bigr) - \tilde{V}\bigl(\{j\},\{s\} \bigm|\xi\bigr)\text{,}
			\end{aligned}
		\end{align}
		where $\indicator $ is the indicator function. 
	The weak inequality follows from
		\begin{align*}
			K_b^z\bigl((K_b^0 \cup{\{j\}},K_s^0),\xi\bigr) \setminus K_b^z(K^0,\xi)
			\subseteq
			K_b^z\bigl((\{j\},\{s\}),\xi\bigr) \setminus K_b^z\bigl((\emptyset,\{s\}),\xi\bigr)
			\text{,}
		\end{align*}
		since, for fixed $\xi$, the set of players who learn about seller~$s$'s deviation via a path through buyer~$j$ is decreasing in the number of other players who initially know of her deviation.
	
	Observe that ${\VS}\bigl(|K_b^0|,|K_s^0|\bigr) = \mathbb{E}_\xi \tilde{V}\bigl(K_b^0,K_s^0 \bigm|\xi\bigr)$. Therefore, since \eqref{eq:Contagion phase mess} holds for almost every~$\xi$, taking the expectation over~$\xi$ yields \eqref{eq:Contagion phase claim}.
\end{proof}

\section{Discussion}

In markets where traders cannot contractually commit to their terms of trade, word-of-mouth communication is viewed to be a powerful incentive: traders may cut ties with those revealed to be defectors, while continuing business with non-defectors. We begin with the premise that traders may not truthfully communicate who is guilty unless they have an incentive to do so. Based on this premise, we find that markets in which traders on only one side have a myopic incentive to shirk can support significantly higher volumes of trade than those in which traders on both sides face moral hazard. The rationale is that traders who lack a myopic incentive to shirk become "guardians" who communicate truthfully to others. Their truthful communication deters traders on the other side of the market from defecting. 

While our model is stylized, these results may help us better understand when ostracism succeeds or fails in practice. Certain situations naturally take the form of a sequential-move game. For example, in financial lending, a lender first decides how much to lend, and a borrower then decides whether to repay. Our results speak to why ostracism, with information about borrowers being shared by lenders, is credible and ubiquitous. Analogously, in the long-distance trade model proposed by \cite{greif:93}, merchants first decide whether to trust agents, and agents later decide to reward or exploit that trust. {Thus, even though \citeauthor{greif:93} does not model players' incentives to report or withhold information, our results imply that merchants would have no incentive to withhold information.}

{More recently, \cite{Bernstein-BeyondRelational} documents a network of relationships among original equipment manufacturers (OEMS) and their suppliers. Within each OEM-supplier relationship, supplier behavior is contractually specified in great detail but OEM behavior is not. Thus, a supplier has no legal recourse if the OEM steals its innovation and then puts production out for bid. Recognizing this problem of one-sided moral hazard, one OEM formed a "Supplier Council" to promote communication among suppliers. Through the lens of our model, we interpret this setting as a seller-first protocol (where OEMs are buyers), and the Council as a communication device that increases the rate of communication among sellers.}

In other contexts, one may envision markets where enforcement intermediaries mitigate incentive issues on one side. {For instance, in supply contracts where quality is not legally enforceable, buyers are often given the right to withhold payment if they deem the quality delivered to be "non-conforming." Similarly, franchising arrangements impose detailed, legally enforceable requirements on the details of franchisees' business operations, but impose few requirements on franchisors \citep{BlairLafontaine-Franchising}.} Such arrangements enable multilateral enforcement because parties who no longer have a myopic incentive to deviate are truthful conduits of information.%
\footnote{Our point complements \cite{acemoglu2015sustaining}, who show how community enforcement can subtly improve enforcement intermediation whereas we focus on the reverse channel.}

By contrast, \cite{bolton2013engineering} discuss how before eBay payments were made through Paypal, both buyers and sellers could deviate, but a two-sided feedback system failed to produce reliable reviews and discipline players. Once it was feasible to structure payments through Paypal, so that buyers no longer needed to be rated, a one-sided feedback system has remained, and such feedback influences sellers' payoffs. Other platforms continue to face issues of two-sided moral hazard. As discussed by \cite{tadelis2016reputation}, Airbnb owners can misrepresent their unit, leave it dirty, etc., and renters too can cheat. In such cases, our theory highlights why players may have strategic reasons not to report deviations they have observed.

Our stylized model omits several  considerations. We study only one motive to conceal information, ignoring costs of communication and the possibility for retribution.  
Moreover, one may view permanent ostracism to take the principle of "ostracizing the guilty, cooperating with the innocent" to a logical extreme. Perhaps after several individuals have been ostracized, it need not be the case that innocent players continue to trade with other innocent players. Our results suggest that combining permanent ostracism with other schemes (e.g., contagion) could be fruitful in settings with two-sided moral hazard, but are unnecessary in those with one-sided moral hazard.

\appendix
\renewcommand{\theequation}{\arabic{equation}}
  \renewcommand{\thesection}{\Alph{section}}
\small

\titlelabel{Appendix~\thetitle\quad}

\section{Definition of Permanent Ostracism}

\titlelabel{\thetitle\quad}

In a permanent ostracism equilibrium, each player~$i$ has a personal state variable, $\omega^i \subset \society$ that lists the players that $i$ deems guilty. Player~$i$'s behavior in each interaction depends on the history in a way that is measurable with respect to $\omega^i$.\footnote{Although $\omega^i$ is a function of player~$i$'s private history, we suppress the history argument except where needed for clarity.} For brevity, we define permanent ostracism only for simultaneous and buyer-first protocols.  

At the start of the game, $\omega^i = \emptyset$.
Under a simultaneous protocol, any player can become guilty. However, under a buyer-first protocol, buyers cannot become guilty (they have "immunity"), because only sellers are subject to moral hazard. We write the set of players with immunity as $\mathcal{I}=\emptyset$ for a simultaneous protocol, and $\mathcal{I} = \buyers$ for a buyer-first protocol.
When a buyer~$b$ and a seller~$s$ meet, and their personal states at the start of the trading stage are $\omega^b$ and $\omega^s$ respectively, then under a simultaneous protocol the buyer should pay $p^*_{bs}(\omega^b)$ and the seller should deliver quality $q^*_{bs}(\omega^s)$. Under a buyer-first protocol, the buyer should pay $p^*_{bs}(\omega^b)$, and then seller should deliver quality $q^*_{bs}(\omega^s)$ if the buyer paid correctly, but deliver quality zero otherwise.
When player~$i$ meets player~$j$ at time~$t$, his personal state updates at the end of each stage of the interaction. We write $\omega^{i-}$ for his state at the start of the stage, and $\omega^{i+}$ for his state at the end of the stage. At the end of the communication stage, after the partners exchange messages $m_i$ and $m_j$, $i$'s state updates from $\omega^{i-}$ to $\omega_i^+ = \bigl( \omega_i^- \cup m_i \cup m_j \bigr) \setminus \mathcal{I}$.
Then, again at the end of the trading stage, the state updates from $\omega^{i-}$ to $\omega^{i+}$ as follows, for each $\ell \in \{i,j\}$:
\begin{itemize}[noitemsep]
    \item \emph{For a simultaneous protocol:} If $\ell \notin \mathcal{I}$ and player~$\ell$ plays any action other than $p^*(\omega^{i-})$ (if player~$\ell$ is the buyer) or $q^*(\omega^{i-})$ (if player~$\ell$ is the seller), then $\ell \in \omega^{i+}$;
    \item \emph{For a buyer-first protocol:} If player $\ell$ is the seller and either (1)~the buyer paid $p^*(\omega^{i-})$ and $\ell$ delivers quality not equal to $q^*(\omega^{i-})$, or (2)~the buyer paid any amount other than $p^*(\omega^{i-})$ and $\ell$ delivers quality not equal to zero, then $\ell \in \omega^{i+}$;
    \item Otherwise $\ell \in \omega^{i-} \iff \ell \in \omega^{i+}$.
\end{itemize}

\begin{definition}\label{Def:PermOstBuyerFirst}
An assessment (a strategy profile and a system of beliefs) is a \emph{permanent ostracism} assessment if there exists a price function $p^*_{bs}: 2^\mathcal{N} \rightarrow \mathbb{R}_+$ and quality function  $q^*_{sb}: 2^\mathcal{N} \rightarrow [0,\bar{q}]$ for each buyer-seller pair~$sb$; each player $i$'s personal state~$\omega^i$ evolves according to the rule given above; and for every player~$i$ and every partner $j \neq i$, if $i$ meets~$j$ at time~$t$, the following are satisfied:
\begin{enumerate}[noitemsep]
    \item In the communication stage, $i$ sends the message $\omega_i \setminus \{i\}$.
    \item{In the trading stage, if the protocol is simultaneous,}
    \begin{enumerate}[noitemsep]
        \item{
             if $\{i,j\} \cap \omega^i = \emptyset$ then $i$ pays $p^*_{ij}(m^i \cup m^j)$ (if $i$ is the buyer) or delivers $q^*_{ij}(m^i \cup m^j)$ (if $i$ is the seller);
        }
        \item{
             if $j \in \omega^i$, then $i$ pays 0 (if $i$ is the buyer) or delivers 0 (if $i$ is the seller).
        }
    \end{enumerate}
    \item{In the trading stage, if the protocol is buyer-first:}
    \begin{enumerate}[noitemsep]
        \item{
             if $i$ is the buyer: $i$ pays $p^*_{ij}(m^i \cup m^j)$ if $j \notin \omega^i$, but pays zero otherwise;
        }
        \item{
             if $i$ is the seller: $i$ delivers $q^*_{ij}(m^i \cup m^j)$ if $i \notin \omega^i$ and $j$ paid $p^*_{ij}(m^i \cup m^j)$, but delivers zero otherwise;
        }
       \end{enumerate}
    \item{\label{POBelief}
    When player~$i$'s state is $\omega^i$ at the start of any communication or trading stage when interacting with player~$j$, player~$i$ assigns probability~1 to the event that $\omega^j \subseteq \omega^i$.}
    
\end{enumerate}
\end{definition}

The requirement on beliefs (\cref{POBelief}) embodies ostracism: as long as player~$i$ has seen no indication---either directly or via messages from other players---that player $k$ may have deviated, $i$~should not believe that $k$ has deviated and caused other players to deem $k$ guilty.

\pagebreak
\begin{singlespace}
{\small
	\addcontentsline{toc}{section}{References}
	\setlength{\bibsep}{.25\baselineskip}
	\bibliographystyle{aer}
	\bibliography{ostbib}
}
\end{singlespace} 
\pagebreak
\section{Supplementary Appendix (Not for Publication)}

\begin{proof}[Proof of \cref{Prop-PermOst}]
  
Suppose towards a contradiction that there is an interaction between players~$i$ and $j$ at which their private histories at the start of the communication stage are $(h_i, h_j)$, their personal states are $(\omega^{i}(h_i),\omega^{j}(h_j))$, and they exchange messages $m^i$ and $m^j$, such that at the start of the trading stage both $i$ and $j$ deem both $i$ and $j$ innocent, and $q^*( m^i \cup m^j ) > \underline{q}$. 

Consider another private history $\hat{h}_i$ that coincides with~$h_i$ except that every player other than $i$ and $j$ has transitioned to being deemed guilty by player~$i$ (so $\omega^i(\hat{h}_i) = \society \setminus \{i,j\}$) after the last interaction in~$h_i$. 
	Suppose player~$j$ communicates first and sends message~$m^j = \omega^j(h_j)$.
	In a permanent ostracism equilibrium, player~$i$ deems player~$j$ innocent, and so should report $m^i = \society \setminus \{i,j\}$ truthfully. Then they should trade at quality $\hat{q} = q^*(\society \setminus \{i,j\})$ and price $\hat{p} = p^*(\society \setminus \{i,j\})$. 
Note that $\hat{q}\leq\underline{q}$ and $\hat{p}\leq\underline{p}$, since they must employ bilateral enforcement in their relationship while permanently ostracizing all other players.
However, if player~$i$ is the seller then a deviation in which he reports~$m^i = \omega^i(h_i)$ rather than~$\omega^i(\hat{h}_i)$ and shirks yields a payoff of
		\begin{align*}
			p^*\bigl(\omega^i(h_i) \cup  \omega^j(h_j)\bigr)
			> \underline{p}
			= \underline{p} - c(\underline{q}) + \int_0^\infty e^{-rt} \lambda (\underline{p} - c(\underline{q})) \,dt
			\text{,}
		\end{align*}
		where the first inequality is by our supposition, the equality is by  definition of~$\underline{p}$ and $\underline{q}$. 
	Similarly, if player~$i$ is the buyer, falsely reporting $\omega^i(h_i)$ and then reneging on payment yields a payoff of
		\begin{align*}
			q^*\bigl(\omega^i(h_i) \cup  \omega^j(h_j)\bigr) 
			> \underline{q}
			= \underline{q} - \underline{p} + \int_0^\infty e^{-rt} \lambda (\underline{q} - \underline{p}) \,dt
				\text{.}
		\end{align*}
	Adding these inequalities yields 
    	\begin{align*}
			p^*\bigl(\omega^i(h_i) \cup  \omega^j(h_j)\bigr) + q^*\bigl(\omega^i(h_i) \cup  \omega^j(h_j)\bigr) 
			&> \underline{p} + \underline{q}
			\\ 
			&= \underline{q} - c(\underline{q}) + \int_0^\infty e^{-rt} \lambda (\underline{q} - c(\underline{q})) \,dt\\
			&\geq \hat{q} - c(\hat{q}) + \int_0^\infty e^{-rt} \lambda (\hat{q} - c(\hat{q})) \,dt 
			\text{,}
		\end{align*}
		where the last inequality follows from $\underline{q} \geq \hat{q}$ and $q-c(q)$ being strictly increasing.
	Therefore, at least one of these deviations is strictly profitable, and so we have reached a contradiction.
\end{proof}

\end{document}